\pdfoutput=1
\documentclass[10pt,twocolumn,twoside]{IEEEtran} 
\usepackage{cite}      % Written by Donald Arseneau
\usepackage{graphicx}  % Written by David Carlisle and Sebastian Rahtz
\usepackage{psfrag}    % Written by Craig Barratt, Michael C. Grant,
\usepackage{subfigure} % Written by Steven Douglas Cochran
\usepackage{url}       % Written by Donald Arseneau
\usepackage{stfloats}  % Written by Sigitas Tolusis
\usepackage{amsmath}   % From the American Mathematical Society
\usepackage{epsfig,graphicx,url,color,bbm, amsthm,amsfonts,dsfont,array,psfrag,stmaryrd,wasysym}

\usepackage{amssymb, amsmath}

\usepackage[T1]{fontenc}
\usepackage{textcomp}
\usepackage{pifont}

\usepackage[ruled,vlined]{algorithm2e}

\usepackage{lineno}

\usepackage{multirow}

\def\mathbi#1{\textbf{\em #1}}

\newcommand{\temps}{t}  %indice de temps
\DeclareMathOperator{\tr}{\textrm{tr}} % trace

\newcommand{\iter}{k}
\newcommand{\iterFct}{i}
\newcommand{\nbFct}{I}
\newcommand{\sizeL}{J}
\newcommand{\rindice}{l}  %indice rank

\DeclareMathOperator{\prox}{prox}

\newcommand{\transpose}{\intercal} % transpose only 
\DeclareMathOperator{\eqdef}{\triangleq} % equality with the meaning "Definition" 

\DeclareMathOperator{\conv}{\star}

\newtheorem{theorem}{Theorem}
\newtheorem{proposition}[theorem]{Proposition}

\newcommand{\defeq}{\triangleq}

\newcommand{\inprod}[1]{\left\langle #1 \right\rangle }
\DeclareMathOperator{\stftFrame}{\Psi}
\DeclareMathOperator{\stftFrameOp}{\psi}

\DeclareMathOperator{\Id}{\mathbf{I}}

\DeclareMathOperator{\compOp}{\mathcal{L}}
\DeclareMathOperator{\compOpBig}{\mathbf{L}}

\DeclareMathOperator{\vect}{vec}

\DeclareMathOperator{\rank}{rank}
\def\LRset{\mathcal{C}_{\rindice}}
\def\LRProj{\mathcal{P}_{\LRset}}

\def\LRMagset{\mathcal{E}_{\rindice}}
\def\LRMagProj{\mathcal{P}_{\LRMagset}}

\newcommand{\nbsamples}{T}
\newcommand{\nbFreqs}{F}
\newcommand{\nbFrames}{Q}
\newcommand{\nbcoeff}{B}

\newcommand{\argmin}{\operatornamewithlimits{argmin}}
\DeclareMathOperator{\weightMtx}{W}
\newcommand{\weightcoeff}{w}

\DeclareMathOperator{\mixop}{\mathcal{A}}

\newcommand{\framebound}{\nu}

\newcommand{\gammaDR}{\gamma}

\def\scvariable{\mathbf{z}}
\def\thvariable{\mathbf{y}}

\def\frstvariableA{\mathbi{s}}
\def\scvariableA{\mathbi{z}}
\def\thvariableA{\mathbi{y}}

\newcommand{\mySigma}{\mbox{\scriptsize \boldmath $\Sigma$}}
\newcommand{\mySigmaEntry}{\mbox{\scriptsize $\Sigma$}} %scalar

\newcommand\revision[1]{{#1}}
\newlength \myfigwidth

\makeatletter %necessary to be able to use the @
\if@twocolumn
\else
\fi

\if@twocolumn
\else
\fi

\newlength \mycelltabwidth
\def\SARABPDN{SSRA}

\def\HPE{PSDNN}
\def\SLRRA{SSLR}

\def\eq_sdmmPb{eq_convex_problem}

\begin{document}

%\linenumbers

\title{\LARGE{Reverberant Audio Source Separation via Sparse and Low-Rank Modeling}}

\author{Simon~Arberet,~Pierre~Vandergheynst % <-this % stops a space
%\author{Pierre~Vandergheynst % <-this % stops a space
%\thanks{}% <-this % stops a space
\thanks{
Copyright (c) 2013 IEEE. Personal use of this material is permitted. However, permission to use this material for any other purposes must be obtained from the IEEE by sending a request to pubs-permissions@ieee.org.

The authors are with the Signal Processing Laboratory, Electrical Engineering Department, \'Ecole Polytechnique F\'ed\'erale de Lausanne (EPFL), Station 11, CH-1015 Lausanne, Switzerland.
\protect\\
(e-mail: simon.arberet@gmail.com)}}

\maketitle

\begin{abstract}

The performance of audio source separation from underdetermined convolutive mixture assuming known mixing filters can be significantly improved
by using an analysis sparse prior optimized by a reweighting $\ell_1$ scheme and a wideband data-fidelity term, as demonstrated by a recent article.
In this letter, we show that the performance can be improved even more significantly by exploiting a low-rank prior on the source spectrograms.
 %i.e. the magnitude of the sources in the time-frequency representation.
We present a new algorithm to estimate the sources based on  i) an analysis sparse prior, ii) a reweighting scheme so as to increase the sparsity, iii)  a wideband data-fidelity term in a constrained form, and iv) a low-rank constraint on the source spectrograms.
Evaluation on reverberant music mixtures shows that the resulting algorithm improves state-of-the-art methods by more than 2 dB of signal-to-distortion ratio.

\end{abstract}

\section{Introduction}

%Most audio recordings
%%such as, CD audio, radio and TV broadcast are 
%can be viewed as mixtures
%of several audio signals (e.g., musical instruments or speech),
%called {\it source signals} or {\it sources}.
%that are usually active simultaneously.
%The sources may have been mixed synthetically with a mixing console
%or by recording a real audio scene using microphones.
An audio recording can be viewed as a mixture 
of several audio signals (e.g., musical instruments or speech), called {\it sources}.
Mathematically, a \textit{convolutive} mixture of $N$ audio sources on $M$ channels can be written as:
\begin{equation}
\label{eq_convolutive_mixture_model}
x_{m}(\temps) = \sum_{n=1}^N  (a_{mn} \conv s_n) (\temps) + e_m(\temps), \quad 1 \leq m \leq M,
\end{equation}
where $s_{n}(\temps) \in \mathbb{R}$ and $x_{m}(\temps) \in \mathbb{R}$ denote sampled time signals of respectively the $n$-th source and the $m$-th mixture
($\temps$ being a discrete time index), $a_{mn}(\temps) \in \mathbb{R}$ denote
the finite (sampled) impulse response of some causal filter, and $\conv$ denotes convolution.

The goal of the \textit{Blind Source Separation (BSS)} problem 
is to estimate the $N$ source signals $s_{n}(\temps)$ ($1 \leq n \leq N$), given the $M$ mixture signals $x_{m}(\temps)$ ($1 \leq m \leq M$).
%standard approach : 
When the number of sources is larger than the number of mixture channels ($N>M$),
the BSS problem is said to be \textit{underdetermined} and is often addressed by sparsity-based approaches \cite{bofill2001underdetermined,Yil04,o2005survey}. %consisting in the following two steps:
%i)
%at the first step the mixing parameters are estimated as in \cite{Yil04,sarberet_demix_journal,arberet2011wideband}, and
%ii) at the second step, the source are estimated. 
%This second step is an underdetermined inverse problem, and as a consequence, some priors such as sparsity, are required to estimate the sources.
%%e.g. using a minimum mean squared error (MMSE) or a Maximum A Posteriori (MAP) estimator 
%%given a sparse source prior and the mixing parameters.

Audio signals are usually not sparse in the time domain, but they are in the time-frequency (TF) domain.
%The estimation of the source coefficients is then usually done in some TF domain using for example the short time Fourier transform (STFT).
%Sparse priors
%There are different ways to model the source sparsity in the TF domain.
Some approaches penalise the source TF coefficients with a $\ell_0$ constraint (binary masking) \cite{Yil04}, or a $\ell_1$ cost \cite{bofill2001underdetermined, kowalski2010beyond}. 
% like binary masking \cite{Yil04} are based on a $\ell_0$ cost.
%is a classical technique which is based on a $\ell_0$ cost.%, assuming disjointness of the sources.
%that, in each TF bin, there is at most one active source, i.e. the source TF representations are disjoint. First a mask is estimated for each source, and then the sources are reconstructed from the TF coefficients given by these masks.
%Other approaches penalise the source TF coefficients with a $\ell_1$ cost \cite{bofill2001underdetermined, kowalski2010beyond}. 
%Minimizing such a sparsity promoting norm, has been shown to be particularly efficient, because first, there are proofs of stable recovery of the sparsest solution (i.e. having the smallest $\ell_0$ cost), and secondly because the $\ell_1$ cost function is convex as opposed to the $\ell_0$ cost which leads to an NP-hard problem. 
%For Long Version
 % instead of the non-convex $\ell_0$ cost function. 
% Basis pursuit denoising (BPDN)  \cite{chen2001atomic} is a very classical approach that use the $\ell_1$ cost to promote sparsity and which has been shown to be particularly efficient. First because there are proofs of stable recovery of the sparsest solution, i.e. having the smallest $\ell_0$ cost,
% secondly because the $\ell_1$ cost function is convex, and thus  the optimization can be solved in a polynomial time as opposed to the $\ell_0$ cost which leads to NP-hard problems.
Another recent approach is the reweighting $\ell_1$ scheme \cite{candes2008enhancing}, which promotes a stronger sparsity assumption than the $\ell_1$ cost, and has recently been shown to outperform $\ell_1$ for source separation by almost 1 dB \cite{arberet2013sparse}.
While \textit{synthesis} sparse priors have been widely used for source modeling, 
\textit{analysis} sparse priors have been used only recently in audio source separation \cite{arberet2013sparse}, and results showed that it improves the separation by about 1 dB in SDR.

%% constraint vs unconstraint
%\revision{The classical BPDN formulation \cite{chen2001atomic} is an unconstrained convex optimization problem, 
%which has an alternative constrained formulation. 
%As for the unconstraint formulation, this constraint formulation %which cannot be solved with a gradient based algorithm, 
%%is however more easy to tune  
%has a regularisation parameter. However, when the noise level is known, the parameter of the constrained formulation is far more easy to set than the unconstrained formulation. 
%Moreover, with this constrained formulation, it is possible to use a reweighing scheme which consists in solving a sequence of (constrained) weighted $\ell_1$ problems,
%and which usually give a sparser solution than the one of the simple $\ell_1$ problem.
%}

%reweighted L1

%% Low rank approaches
%Sparsity modelling is based on the idea that, in a high-dimensional vector space, the data has a low-dimensional structure.
%Low-rank modelling is based on a similar idea but for matrix spaces instead of vector spaces. 
%It assumes that a matrix can be well approximated by its low-rank approximation.
Low-rank modeling, which can be traced back from Eckart \cite{eckart1936approximation} has been widely exploited in problems such as matrix completion \cite{candes2009exact} and robust PCA \cite{wright2009robust}. 
The idea of modeling the source spectrograms  (i.e. the magnitude of the source TF coefficients) 
with a low-rank matrix has not been used directly, but indirectly via the non-negative matrix factorization (NMF)  \cite{Ozerov2010,5605570} which also assumes the non-negativity of the factors. While this idea has been quite successful in audio BSS, it remains that the NMF approximation has some important limitations: its solution is non-unique %, an optimization algorithm with a good initialisation is required to compute a solution,  
and it converges but only to a fix point and very slowly.
However, without these non-negativity constraints, the low-rank approximation, in the least squares sense, is unique and has a closed form solution, which can be computed via a singular value decomposition (SVD).

In this article, we focus on addressing the source estimation task, \revision{i.e. the second stage of a typical BSS approach, }
assuming that the mixing filters $ a_{mn}$ are known.
%We propose a novel algorithm, for convolutive source separation which is based on the assumption that the sources are sparse and have a low-rank spectrogram in the TF domain.
The main contribution of this paper is to: i) introduce, in addition to a sparsity assumption, a low-rank model of the source spectrograms, i.e. we assume that the magnitude (and not the phase) of the short-time Fourier representation of each source is low-rank, and ii) derive an optimization algorithm based on a proximal splitting scheme \cite{combettes2011proximal} so as to estimate the sources.
This algorithm also incorporates three ingredients, which were recently introduced in audio BSS \cite{arberet2013sparse}:
 i) an analysis sparsity prior, %which is fundamentally different than the synthesis prior when the analysis operator is a redundant frame (such as a redundant STFT), 
 ii) a reweighting $\ell_1$ scheme, %that mimics the $\ell_0$ minimization so as to promote a stronger sparsity assumption than the $\ell_1$ cost,  
 and iii) a wideband data fitting constraint. %and thus first avoid the narrowband approximation, and secondly offers a strong fidelity term (in a new constrained formulation) without the need to fix a regularization parameter (in a standard regularized formulation). 
 %1) Analysis prior: more flexible (can mix dif STFT frames,....) and more efficient 2) cst BPDN (no tuning of lambda), 3) reweighting scheme for promoting L0.

The organization of the remainder of the paper is the following. 
We introduce, in section \ref{sec_notations}, our notations, in section \ref{sec_pbform}, the optimization problem we want to solve. In section \ref{sec_optim}, we discuss convex optimization approaches and introduce our algorithm, and in section \ref{sec_expe}, we provide numerical results.

\section{Notations\label{sec_notations}}

\subsection{The convolutive mixture model in operator form\label{sec_convmodel}}

The mixture model \eqref{eq_convolutive_mixture_model} can be written as:
\begin{equation}\label{eq_convolutive_mixture_model_op}
\mathbf{x} =  \mixop(\mathbf{s}) + \mathbf{e}.
\end{equation}
where $\mathbf{x} \in \mathbb{R}^{M \times \nbsamples}$ is the matrix of the mixture composed of the $x_m(t)$ entries, i.e. $ \mathbf{x} = [x_m(t)]_{m=1,t=1}^{M,\nbsamples}$, %with $1 \leq t \leq \nbsamples$, 
$\nbsamples$ being the number of samples. Similarly $\mathbf{s} \in \mathbb{R}^{N \times \nbsamples}$ is the matrix of sources composed of the $s_n(t)$ entries,
$\mathbf{e} \in \mathbb{R}^{M \times \nbsamples}$ is the matrix of the noise composed of the $e_m(t)$ entries,
and 
%$\mathbf{A} \in \mathbb{R}^{M \times N \times \filterLength}$ is the three dimensional array of the mixing filters composed of the $a_{mn}(t)$ entries.
$\mixop : \mathbb{R}^{N \times \nbsamples} \to  \mathbb{R}^{M \times \nbsamples}$ is the discrete linear operator
defined by $$[ \mixop(\mathbf{s}) ]_{m, \temps} =  \sum_{n=1}^N  (a_{mn} \conv s_n) (\temps).$$  
%that apply the convolutive mixing process \eqref{eq_convolutive_mixture_model} on the sources $\mathbf{s}$.
The adjoint operator $\mixop^*: \mathbb{R}^{M \times \nbsamples} \to \mathbb{R}^{N \times \nbsamples}$ of $\mixop$ 
is obtained by applying  the convolution mixing process with the adjoint filters 
$a^*_{nm}(\temps) \defeq a_{mn}(-\temps), \forall \temps$ instead of $a_{mn}$, that is:
$[ \mixop^{*}(\mathbf{x}) ]_{n, \temps} =  \sum_{m=1}^M  (a^{*}_{nm} \conv x_m) (\temps)$.

%Note that Eq. \eqref{eq_convolutive_mixture_model_op} can be written in the following matrix form:
%$$\mathbf{x}_{\rm vec} =  \mathbf{A} \mathbf{s}_{\rm vec} + \mathbf{e}_{\rm vec},$$ 
%where $\mathbf{x}_{\rm vec} \in  \mathbb{R}^{M \nbsamples}$, $\mathbf{s}_{\rm vec} \in  \mathbb{R}^{N \nbsamples}$ and $\mathbf{e}_{\rm vec} \in  \mathbb{R}^{M \nbsamples}$ are  
%the unfolded vectors of the matrices $\mathbf{x}$, $\mathbf{s}$ and $\mathbf{e}$, respectively, and 
% $\mathbf{A}$ is a matrix of size $M \nbsamples \times N \nbsamples$
% composed of $M \times N$ Toeplitz blocks $A_{mn}$ of size $\nbsamples \times \nbsamples$.

\subsection{Time-frequency transform}

%Underdetermined source separation is an ill-posed inverse problem, which needs additional assumptions to be solved.
As stated in the introduction, a powerful assumption is the sparsity of the audio sources in the TF domain. 
A popular TF representation is obtained via the short time Fourier transform (STFT).
%Audio signals are known to be sparse in the TF domain, and a popular TF representation is obtained via the STFT.

The monochannel STFT operator $\stftFrameOp: \mathbb{R}^{\nbsamples} \to \mathbb{C}^{\nbFrames \times \nbFreqs}$ transforms a monochannel signal $\mathbf{s}_n$ of length $\nbsamples$, into a matrix $\stftFrameOp({\mathbf{s}_n}) = [\hat{\mathbf{s}}_n(q L / R,f)]_{q=1,f=1}^{\nbFrames,\nbFreqs} \in \mathbb{C}^{\nbFrames \times \nbFreqs}$ of TF coefficients $\hat{\mathbf{s}}_n(t,f)$, with $t = q L / R $,  $L$ being the window size, $R$ the redundancy ratio, $q$ and $f$, the time frame and frequency index, respectively.
Let us also define the multichannel STFT operator $\stftFrame \in \mathbb{C}^{\nbsamples \times \nbcoeff}$ that transforms a multichannel signal 
$\mathbf{s}$ of length $\nbsamples$, into a matrix $\tilde{\mathbf{s}} \in \mathbb{C}^{N \times \nbcoeff}$ %of $\nbcoeff$ time-frequency coefficients per channel:
populated by the $\nbcoeff = \nbFrames \nbFreqs$ TF column vectors $\hat{\mathbf{s}}(t,f) \in \mathbb{C}^{N}$.
Thus $\tilde{\mathbf{s}} = \mathbf{s}  \stftFrame$, 
%\begin{equation}
%\tilde{\mathbf{s}} = \mathbf{s}  \stftFrame, \nonumber
%\end{equation}
%frame analysis operator $\stftFrame$.
and the ISTFT is obtained by applying the adjoint operator $\stftFrame^* \in \mathbb{C}^{\nbcoeff \times \nbsamples}$ on the STFT coefficients $\tilde{\mathbf{s}}$, i.e. 
$\mathbf{s} = \tilde{\mathbf{s}}   \stftFrame^*.$
With these notations, it is clear that $\mathbf{s}_n  \stftFrame = \vect(\stftFrameOp({\mathbf{s}_n}))$, where $\vect()$ is the vec operator
which maps a matrix into a vector by stacking its columns.
Let also define the \textit{source spectrogram} of source $\mathbf{s}_n$ as $|\stftFrameOp(\mathbf{s}_n)| \in \mathbb{R}_+^{\nbFrames \times \nbFreqs}$, 
where $|\cdot|$ is the element wise absolute value.

\section{Problem formulation\label{sec_pbform}}

In order to estimate the sources from the mixture, we formulate an optimization problem composed of three terms.
%The approach we propose in this paper to estimate the sources from the mixture is based on the resolution of an optimisation problem with three terms.
First, as we want our convolutive mixture model \eqref{eq_convolutive_mixture_model_op} to match the observations, we impose the reconstruction error $\|\mathbf{x}-  \mixop( \mathbf{s}) \|_2$ to be small and bounded by $\epsilon$. 
Secondly, we assume an analysis sparse prior of the source TF representation, and thus we would like to minimize the $\ell_0$ norm 
$\| \mathbf{s} \stftFrame \|_0$. Finally we assume that the rank of 
each source spectrogram $|\stftFrameOp(\mathbf{s}_n)|$ is bounded by a small integer $\rindice$.
%Finally, the problem we would like to solve can be written as:
%\begin{align}\label{eq_L0Pb}
%\argmin_{\mathbf{s} \in \mathbb{R}^{N \times T}}\ & 
%\| \mathbf{s} \stftFrame  \|_{0} \nonumber \\
%\text{subject to}\ & \|\mathbf{x}-  \mixop ( \mathbf{s}) \|_2 \leq \epsilon,\nonumber \\
%& \rank (|\stftFrameOp(\mathbf{s}_n)|) \leq \rindice, \quad n=1,\ldots,N.
%\end{align}

This problem is NP because of the $\ell_0$ norm and thus cumbersome for a problem of our size. 
However, the $\ell_0$ norm can be replaced by a $\ell_1$ norm, or for a sparser solution, by a sequence of weighted $\ell_1$ minimizations $\| \mathbf{s} \stftFrame  \|_{\weightMtx,1} $ where $\weightMtx \in \mathbb{R}_+^{N \times \nbcoeff}$ is a matrix with positive entries $\weightcoeff_{ij}$, and
$\| \mathbf{z}  \|_{\weightMtx,1} \defeq \sum_{i,j} \weightcoeff_{ij} |z_{ij}|$ is the weighted $\ell_1$ norm \cite{candes2008enhancing}. 
Finally, the problem we want to solve, replacing the $\ell_0$ norm with the weighting $\ell_1$ norm is:
\begin{align}\label{eq_reweightedL1Pb}
\argmin_{\mathbf{s} \in \mathbb{R}^{N \times T}}\ & 
%\| \weightMtx \circ (\mathbf{s} \stftFrame)  \|_1 \nonumber \\
\| \mathbf{s} \stftFrame  \|_{\weightMtx,1} \nonumber \\
\text{subject to}\ & \|\mathbf{x}-  \mixop ( \mathbf{s}) \|_2 \leq \epsilon,\nonumber \\
& \rank (|\stftFrameOp(\mathbf{s}_n)|) \leq \rindice, \quad n=1,\ldots,N.
\end{align}

\section{Optimization Algorithms\label{sec_optim}}

In order to estimate the sources, an optimization algorithm called \SLRRA\ is derived.
This (meta-)algorithm solves a sequence of optimization subproblems, each of which involves finding the solution of
problem  \eqref{eq_reweightedL1Pb}. 
%we derive in this section an optimization algorithm called \SLRRA\ to find an approximate solution of problem \eqref{eq_L0Pb} via a $\ell_1$ reweighted method 
%similar as \SARABPDN\ \cite{arberet2013sparse}. 
%This method solves a sequence of optimization subproblems, each of which solves the weighted $\ell_1$ problem  \eqref{eq_reweightedL1Pb}. 
%First, we explain in section \ref{sec_SSRA} the \SARABPDN\ algorithm and how we adapt it to our problem, and then, in section \ref{convexOptim}, we explain how we can find a solution of problem \eqref{eq_reweightedL1Pb}.

\subsection{The \SARABPDN\ and \SLRRA\ algorithms\label{sec_SSRA}}

The SSRA algorithm \cite{arberet2013sparse} is an iterative procedure 
which consists in computing, at each iteration $\iter$, the solution $\mathbf{s}^{(\iter)}$ of  a weighted $\ell_1$ problem, for a given weight matrix
$\weightMtx^{(\iter)}$, and then re-estimating $\weightMtx$ such that the weights $\weightMtx^{(\iter+1)}$ are essentially the inverse of the value of the solution $\mathbf{s}^{(\iter)}$ of the current problem.
This reweighing scheme is a classical procedure \cite{candes2008enhancing,carrilloSACS,arberet2013sparse} which has been proved to approach the $\ell_0$ norm minimization.
In this paper we are using the same reweighting approach as \SARABPDN, but with subproblem \eqref{eq_reweightedL1Pb} instead of 
the weighted $\ell_1$ problem of  \cite{arberet2013sparse} which is essentially the same as problem \eqref{eq_reweightedL1Pb} but without the low-rank constraints.
We call \SLRRA\ the resulting procedure.

\subsection{Convex optimization algorithms\label{convexOptim}}

At each iteration of the reweighing approach described in section \ref{sec_SSRA}, the solution of problem \eqref{eq_reweightedL1Pb} has to be computed.
%Solving problem \eqref{eq_reweightedL1Pb} consists in minimizing a non-smooth convex function under an $\ell_2$-ball constraint and low-rank matrix constraints.
In order to compute the solution of this problem, we rely on the framework of proximal splitting methods \cite{combettes2011proximal},
which are efficient convex optimization algorithms that can deal with non-smooth functions and multiple constraints.
% and which are particularly well suited for large scale problems.
While in Problem \eqref{eq_reweightedL1Pb}, the $\ell_2$-ball is a convex set, the set of low-rank matrices is non-convex. 
However, despite any convergence guaranty in general, using non-convex set constraints in proximal splitting methods can lead to efficient algorithms in practice when the projection can be computed exactly \cite{boyd2011distributed}.

%Hence, it is not possible to use conventional smooth optimization techniques based on the gradient. 
%However we can use proximal optimization methods \cite{combettes2011proximal} that are 
%efficient convex optimization algorithms that can deal with non-smooth functions and which are particularly well suited for large scale problems.

We first introduce the general framework of \textit{proximal splitting methods}.
%We then derive the proximity operators involved in our optimization problem \eqref{eq_reweightedL1Pb}, which defined the elementary operations that are required to fit problem \eqref{eq_reweightedL1Pb} into the general \textit{proximal splitting} framework, 
Then we describe the PSDMM algorithm (Algorithm \ref{algo_Precond_SDMM}) which is a well-adapted algorithm to solve optimization problems involving an arbitrary number of non-smooth functions, and more particularly problem \eqref{eq_reweightedL1Pb}.
% in terms of speed and scalability of the techniques to very high dimensions.

\subsubsection{Proximal splitting methods}
As we will see in section \ref{sec_PSDMM}, proximal splitting methods can solve optimization problems of the form:
\begin{equation}\label{eq_convex_problem}
\argmin_{\mathbf{s} \in \mathbb{R}^{N \times \nbsamples}} \sum_{\iterFct=1}^{\nbFct} f_{\iterFct}(\compOp_i(\mathbf{s})),
\end{equation}
%or 
%\begin{equation}\label{eq_convex_problem}
%\argmin_{\mathbf{z} \in \mathbb{R}^{\nbentries}} f_1(\mathbf{z}) +f_2(\mixop(\mathbf{z}))+f_3(\mathbf{z})
%\end{equation}
where $f_{\iterFct}$, are convex functions from $\mathbb{R}^{\sizeL_{\iterFct}}$ to $\mathbb{R}$ and $\compOp_{\iterFct}: \mathbb{R}^{N \times \nbsamples} \to  \mathbb{R}^{\sizeL_{\iterFct}}$ are bounded linear operators.
Note that any convex constraint $C$ on $\mathbf{s}$ can be incorporated in this formulation via the indicator function $i_C(\cdot)$, 
 where $C$ represents the constraint set, and $i_C(\mathbf{s}) = 0$ if $\mathbf{s} \in C$, and $+ \infty$ otherwise.

%Note that any convex constrained problem can be formulated
%as an unconstrained problem by using the indicator function $i_C(\cdot)$
%of the convex constraint set $C$ as one of the functions in \eqref{eq_convex_problem}, e.g. $f_2(\mathbf{z})=i_C(\mathbf{z})$ where $C$ represents the constraint set,
%and $i_C(\mathbf{z}) = 0$ if $\mathbf{z} \in C$, and $+ \infty$ otherwise.

Problem \eqref{eq_reweightedL1Pb} can be seen as a particular instance of problem \eqref{eq_convex_problem} with three functions $f_1$, $f_2$, $f_3$, and 
with $\compOp_1 = \compOp_3 =\Id$, $\compOp_2 = \mixop$, 
$f_1(\mathbf{s} ) = \| \mathbf{s} \stftFrame  \|_{\weightMtx ,1}$, $f_2(\mixop(\mathbf{s})) = i_{\mathcal{B}_{\ell_2}^{\epsilon}}( \mixop(\mathbf{s}))$, where 
%$\mathcal{B}_{\ell_2}^{\epsilon}$
$\mathcal{B}_{\ell_2}^{\epsilon} = \{\mathbf{s}  \in \mathbb{R}^{N \times \nbsamples}:\ \| \mathbf{s} -  \mathbf{x} \|_2 \leq\epsilon \}$, and 
 $f_3(\mathbf{s}) = i_{\mathcal{R}^{\rindice}}(\mathbf{s})$, where 
 %$ \mathcal{R}^{\rindice} =   \{\mathbf{s}  \in \mathbb{R}^{N \times \nbsamples} :\  \rank (|\mathbf{s} \stftFrame |) \leq \rindice \}$.
$\mathcal{R}^{\rindice} =   \{\mathbf{s}  \in \mathbb{R}^{N \times \nbsamples}:\   1 \leq n \leq N,\ \rank (|\stftFrameOp(\mathbf{s}_n)|) \leq \rindice \}$.
Note that $f_1(\mathbf{s} )$ and $f_2(\mixop(\mathbf{s}))$ are convex, but $f_3(\mathbf{s})$ is not convex because $\mathcal{R}^{\rindice}$ is a non-convex set.

The key concept in proximal splitting methods is the use of the proximity
operator $\prox_{f_i}$ of a function $f_i$ defined as:
 \begin{equation}
\prox_{f_i}(\mathbf{z}) \ \eqdef\  \argmin_{\mathbf{y} \in \mathbb{R}^{\sizeL_{\iterFct}} } f_i(\mathbf{y}) + \frac{1}{2} \| \mathbf{z} - \mathbf{y} \|_2^2, 
\end{equation}
\noindent which is a natural extension of the notion of a projection.
This definition extends naturally for some matrices $\mathbf{z}$ and $\mathbf{y}$, by replacing the $\ell_2$ norm with the Frobenius norm.
%For example, the proximal operator of the $\ell_1$ norm is the soft-thresolding
%operator, and the proximal operator of the indicator function of a
%constraint is simply the projection operator onto the constraint set.
%Proximal splitting methods proceed by splitting the contribution of
%the functions $f_1$, $f_2$ individually so as to yield an easily implementable
%algorithm. They are called proximal because each nonsmooth
%function in \eqref{eq_convex_problem} is involved via its proximity operator. In
%essence, the 
Solution to \eqref{eq_convex_problem} is reached iteratively by successive application
of the proximity operator associated with each function $f_{\iterFct}$.
See \cite{combettes2011proximal} for a review of proximal splitting
methods and their applications in signal and image processing.

We derive in the appendix the proximity operators of functions
$f_1(\mathbf{s} ) = \|  \mathbf{s} \stftFrame  \|_{\weightMtx,1}$, $f_2(\mathbf{s}) = i_{\mathcal{B}_{\ell_2}^{\epsilon}}(\mathbf{s})$ and 
 $f_3(\mathbf{s}) = i_{\mathcal{R}^{\rindice}}(\mathbf{s})$ 
involved in optimization problem \eqref{eq_reweightedL1Pb}, 
%The proximal operator of function $f_3(\mathbf{s}) = i_{\mathcal{R}^{\rindice}}(\mathbf{s})$ is calculated in section \ref{sec_lowrankProjection}.
%We now assume that we can compute the proximity operators of each of the functions involved in our optimization problem, and we derive in the following sections the framework to solve problem \eqref{eq_convex_problem}, 
and we derive in the following sub-sections the optimization framework to solve problem \eqref{eq_convex_problem}.

%However we can use a method such as the \textit{Douglas-Rachford} method (DR) \cite{combettes2011proximal}.
%This requires to define the proximal operator of the $\ell_1$ norm which is classical soft thresholding, and the $\ell_2$-ball projection.

%\subsubsection{Chambolle-Pock Algorithm}
%
%Antonin Chambolle and Thomas Pock \cite{chambolle2011first} proposed a primal-dual scheme to solve the convex optimizitaion problem:
%
%\begin{equation}\label{eq_cp_convex_problem}
%\argmin_{\mathbf{z} \in \mathcal{H} } F(\compOpBig(\mathbf{z})) + G(\mathbf{z})
%\end{equation}
%\noindent where $F$ and $G$ are convex functions.
%
%
% \begin{algorithm}  [t!]        \label{algo_CP}           % enter the algorithm environment
%\text{Initialize:} $\iter=0$, $\scvariable^{(0)} \in \mathbb{R}^{N \times \nbsamples}$, $\alpha_{\iter} \in (0,2)$, $\gammaDR > 0$.\\
%\Repeat{}{
%$\thvariable^{(\iter+1)} = \prox_{\gamma F^*} (\thvariable^{(\iter)} + \gamma \compOpBig(\bar{\frstvariable}^{(\iter)}) )$\\
%$\frstvariable^{(\iter+1)} = \prox_{\tau G}(\frstvariable^{(\iter)} - \tau \compOpBig^* (\thvariable^{(\iter+1)})) $\\
%$\bar{\frstvariable}^{(\iter+1)} = \frstvariable^{(\iter+1)} + \theta (\frstvariable^{(\iter+1)} -\frstvariable^{(\iter)} )$\\
%$\iter = \iter+1$.\\
%}( convergence)
%\Return $\frstvariable^{(\iter)}$ \\
%\caption{Chambolle-Pock algorithm} 
%\end{algorithm}

 % &= \| Y \|_F^2 + \| |X| \|_F^2 - 2 \tr((|X|)^{\tc} (|Y|e^{i (\angle Y-\angle X)})) \\

% &= \| Y \|_F^2 + \| |X| \|_F^2 - 2 (  \vect(|X|)\circ \vect(|Y|) \circ \vect e^{i (\angleY - \angle X)}   \\

\subsubsection{ADMM Algorithm}

The Alternating Direction Method of Multipliers (ADMM) \cite{combettes2011proximal} is a well suited algorithm 
to solve large-scale convex optimization of the form:

\begin{equation}\label{eq_cp_convex_problem}
\argmin_{\mathbi{s} \in \mathcal{H} } F(\compOpBig(\mathbi{s})) + G(\mathbi{s}), 
\end{equation}
\noindent where  $F: \mathcal{G} \to \left]- \infty,+\infty \right]$ and  $G: \mathcal{H} \to \left]- \infty,+\infty \right]$ are proper, convex, lowersemicontinuous (l.s.c.) functions, 
$\mathcal{H}$ and  $\mathcal{G}$ being finite-dimensional real vector spaces equipped with an inner
product $\inprod{\cdot,\cdot}$, and a norm $\| \cdot\| = \inprod{\cdot,\cdot}^{\frac{1}{2}}$. The map $\compOpBig:  \mathcal{H} \to \mathcal{G} $ is a continuous linear operator with induced norm: $\|\compOpBig \| = \max \{\|\compOpBig(\mathbi{s}) \|:  \mathbi{s} \in \mathcal{H}\ \text{with}\ \|\compOpBig (\mathbi{s}) \| \leq 1 \}.$
If $\compOpBig$ is injective, the ADMM algorithm described in Algorithm \ref{algo_ADMM} converges to a solution of \eqref{eq_cp_convex_problem},
where we denoted by $\prox_{G}^{\compOpBig}$ the operator which maps a point $ \thvariableA \in \mathcal{G}$ to the
unique minimizer of $\mathbi{s} \mapsto  G(\mathbi{s}) + \frac{1}{2} \| \compOpBig (\mathbi{s}) - \thvariableA \|_2^2.$

 \begin{algorithm}  [t!]        \label{algo_ADMM}           % enter the algorithm environment
\text{Initialize:} $\iter=0$, $\thvariableA^{(0)} \in  \mathcal{G}$, $\scvariableA^{(0)} \in  \mathcal{G}$, $\gamma > 0$.\\
\Repeat{}{
$\frstvariableA^{(\iter+1)} = \prox_{\gamma G}^{\compOpBig} (\thvariableA^{(\iter)} - \scvariableA^{(\iter)})$\\
$\thvariableA^{(\iter+1)} = \prox_{\gamma F}(\compOpBig (\frstvariableA^{(\iter+1)}) + \scvariableA^{(\iter)})$\\
$\scvariableA^{(\iter+1)} = \scvariableA^{(\iter)} + \compOpBig (\frstvariableA^{(\iter+1)}) - \thvariableA^{(\iter+1)}$\\
$\iter = \iter+1$.\\
}( convergence)
\Return $\frstvariableA^{(\iter)}$ \\
\caption{ADMM algorithm} 
\end{algorithm}

%The minimization
%\begin{align}
%\frstvariableA^{(\iter+1)}  & = 
%\prox_{\gamma G}^{\compOpBig} (\thvariableA^{(\iter)} - \scvariableA^{(\iter)})\\
% & \eqdef \argmin_{\frstvariableA \in \mathcal{H}}  \gamma G(\frstvariableA) + \frac{1}{2} \|\compOpBig (\frstvariableA) - (\thvariableA^{(\iter)}-\scvariableA^{(\iter)}) \|^2 \\
% & = \argmin_{\frstvariableA \in \mathcal{H}}  G(\frstvariableA) + \frac{1}{\gamma}  \inprod{\scvariableA^{(\iter)},\compOpBig (\frstvariableA)} + \frac{1}{2 \gamma} \|\compOpBig (\frstvariableA) - \thvariableA^{(\iter)} \|^2 \label{eq_xupdateADMM}
% \end{align}
%is a  least squares problem including the linear operator $\compOpBig$ which computation necessitates inner iterations.
%Antonin Chambolle and Thomas Pock \cite{chambolle2011first} proposed a trick to precondition this step by adding, in the minimization \eqref{eq_xupdateADMM}, the  
%the following term:
%$\frac{1}{2} \inprod{(\frac{1}{\tau} - \frac{1}{\gamma} \compOpBig\compOpBig^*)(\frstvariableA-\frstvariableA^{(\iter)}), \frstvariableA-\frstvariableA^{(\iter)}}$,
%with $\tau < \frac{\gamma}{\| \compOpBig \|^2}$.  Adding this term to the $\frstvariableA$-update  step \eqref{eq_xupdateADMM} yields the following update:
%$$\frstvariableA^{(\iter+1)} = \prox_{\tau G} (\frstvariableA^{(\iter)} -\tau \compOpBig^* ( \bar{\frstvariableA}^{(\iter)}))$$
%with $\bar{\frstvariableA}^{(\iter)} = \frac{1}{\gamma}(2 \scvariableA^{(\iter)} -\scvariableA^{(\iter-1)}),$

Minimization $ \frstvariableA^{(\iter+1)}   = \prox_{\gamma G}^{\compOpBig} (\thvariableA^{(\iter)} - \scvariableA^{(\iter)})$
 is a  least squares problem including the linear operator $\compOpBig$ which computation necessitates inner iterations.
Antonin Chambolle and Thomas Pock \cite{chambolle2011first} proposed a trick to precondition this step.
Using their preconditioner (see section \ref{sec_ChambollePock} in the Appendix), this minimization can be replaced by a simple prox computation, 
yielding the preconditioned ADMM algorithm also known as Chambolle-Pock Algorithm. % described in Algorithm \ref{algo_precond_ADMM}.
Interestingly, the convergence of this algorithm has been proved \cite{chambolle2011first} for a general (not necessarily injective) bounded linear operator $\compOpBig$. 

% \begin{algorithm}  [t!]        \label{algo_precond_ADMM}           % enter the algorithm environment
%\text{Initialize:} $\iter=0$, $\frstvariableA^{(0)} \in  \mathcal{H}$, $\scvariableA^{(0)} \in  \mathcal{G}$, $\gamma > 0$, $\tau < \frac{\gamma}{\| \compOpBig \|^2}$\\
%\Repeat{}{
%$\thvariableA^{(\iter+1)} = \prox_{\gamma F}(\compOpBig (\frstvariableA^{(\iter)}) + \scvariableA^{(\iter)})$\\
%$\scvariableA^{(\iter+1)} = \scvariableA^{(\iter)} + \compOpBig (\frstvariableA^{(\iter)}) - \thvariableA^{(\iter+1)}$\\
%$\frstvariableA^{(\iter+1)} = \prox_{\tau G} (\frstvariableA^{(\iter)} -\tau \compOpBig^* ( \bar{\frstvariableA}^{(\iter)}))$\\
%with $\bar{\frstvariableA}^{(\iter)} = \frac{1}{\gamma}(2 \scvariableA^{(\iter+1)} -\scvariableA^{(\iter)})$\\
%$\iter = \iter+1$.\\
%}( convergence)
%\Return $\frstvariableA^{(\iter)}$ \\
%\caption{Preconditioned ADMM algorithm} 
%\end{algorithm}

\subsubsection{ Preconditioned SDMM (\HPE) Algorithm\label{sec_PSDMM}}

%In order to be able to take into account multiple regularizers, we aim at minimizing optimimization problem \eqref{eq_convex_problem}.
%%sol = argmin SUM(f_i(L_i*x))for x belong to R^N and L_i linear operators
%% [1] Combettes and J-C. Pesquet, "Proximal Splitting Methods 
%% in Signal Processing", IEEE Journal
%\begin{equation}
%\argmin_{\mathbf{s} \in \mathbb{R}^{N \times \nbsamples}} \sum_{\iterFct=1}^{\nbFct} f_{\iterFct}(\compOp_{\iterFct} (\mathbf{s})),
%%\label{eq_sdmmPb}
%\end{equation}
%where  $\compOp_{\iterFct}: \mathbb{R}^{N \times \nbsamples} \to  \mathbb{R}^{\sizeL_{\iterFct}}$ are bounded linear operators, 
%and where $ f_{\iterFct}:  \mathbb{R}^{\sizeL_{\iterFct}} \to \left]- \infty,+\infty \right]$ are l.s.c. convex functions.  
In a similar way as in \cite{combettes2011proximal}, problem \eqref{eq_convex_problem} can be formulated as a particular case of problem \eqref{eq_cp_convex_problem} in the 
$\nbFct$-fold product space $\mathcal{H} = \mathbb{R}^{N \times \nbsamples} \times \ldots \times   \mathbb{R}^{N \times \nbsamples}$,
with $\mathcal{G} = \mathbb{R}^{\sizeL_{1}} \times \ldots \times   \mathbb{R}^{\sizeL_{\nbFct}}$.
We denote $\mathbi{s} = (\mathbi{s}_1,\ldots,\mathbi{s}_{\nbFct})$ a generic element of $\mathcal{H}$, 
and $\mathbi{z} = (\mathbi{z}_1,\ldots,\mathbi{z}_{\nbFct})$ a generic element of $\mathcal{G}$.
Then we define 
$\compOpBig: \mathcal{H} \to \mathcal{G}$ by $\compOpBig(\mathbi{s}) = (\compOp_{1}(\mathbi{s}_1),\ldots,\compOp_{\nbFct}(\mathbi{s}_{\nbFct}))$,
$F(\mathbi{z}) =  \sum_{\iterFct=1}^{\nbFct} f_{\iterFct} (\mathbi{z}_{\iterFct})$, and $G(\mathbi{s}) = i_D(\mathbi{s})$ where,  $ i_D(\cdot)$ the indicator function of the convex set $D=\{(\mathbf{s},\ldots,\mathbf{s}) \in \mathcal{H}:\ \mathbf{s} \in \mathbb{R}^{N \times \nbsamples}  \}$.
By deriving algorithm \ref{algo_ADMM} with this parametrization and the Chambolle-Pock preconditioner, we obtain algorithm \ref{algo_Precond_SDMM}, denoted \HPE.
%A solution of problem \eqref{eq_sdmmPb} can then be optained by Algorithm \ref{algo_Precond_SDMM}, which is obtained by deriving 
%Algorithm \ref{algo_precond_ADMM} on problem \eqref{eq_sdmmPb} with .

 \begin{algorithm}  [t!]        \label{algo_Precond_SDMM}           % enter the algorithm environment
\text{Initialize:} $\iter=0$, $\mathbf{s}^{(0)} \in \mathbb{R}^{N \times \nbsamples}$, 
for $\iterFct=1,\ldots, \nbFct$, 
$\scvariable^{(\iterFct,0)} \in \mathbb{R}^{\sizeL_{\iterFct}}$, $\gammaDR > 0$, $\tau < \gammaDR /\|\compOpBig \|^2 $\\
\Repeat{}{

\For{$\iterFct\leftarrow 1$ \KwTo $\nbFct$}{
%$\mathbf{u}^{(\iterFct,\iter)} = \compOp_{\iterFct} (\mathbf{s}^{(\iter)})$\\
$\thvariable^{(\iterFct,\iter+1)} = \prox_{\gamma f_{\iterFct}}( \compOp_{\iterFct} (\mathbf{s}^{(\iter)}) +  \scvariable^{(\iterFct,\iter)})$\\
$\scvariable^{(\iterFct,\iter+1)} =  \scvariable^{(\iterFct,\iter)} +  \compOp_{\iterFct} (\mathbf{s}^{(\iter)}) - \thvariable^{(\iterFct,\iter+1)}$\\
}

$\mathbf{s}^{(\iter+1)} = \mathbf{s}^{(\iter)} -  \frac{\tau}{\gammaDR \nbFct } \sum_{\iterFct=1}^{\nbFct} \compOp_{\iterFct}^* (2 \scvariable^{(\iterFct,\iter+1)}  - \scvariable^{(\iterFct,\iter)})$

$\iter = \iter+1$.\\
}( convergence)
\Return $\mathbf{s}^{(\iter)}$ \\
\caption{PSDMM: Preconditioned SDMM algorithm} 
\end{algorithm}

%While the DR algorithm converges when the number of iterations tends to infinity, we have to choose a stopping criterion.
%We chose to stop the algorithm when the relative change of the objective value between two successive estimates is less than a given value $\stopthresh$,
%i.e. $ \left| f_1(\mathbf{s}^{(\iter)}) - f_1(\mathbf{s}^{(\iter-1)}) \right| / f_1( \mathbf{s}^{(\iter)})
% < \stopthresh $, or when the number of iterations is greater than a given value $\maxIter$.
%In our experiments, we fixed $\stopthresh=0.01$, and $\maxIter=200$.

%\input{lowrankProjection}

\section{Experiments \label{sec_expe}}

We evaluated our \SLRRA\ algorithm with state-of-the-art methods over convolutive mixtures of music sources.
For all the experiments, the test signals are sampled at $11$ kHz and we use a STFT with cosine windows. 

\subsection{Experimental protocol}

%We used the same experimental protocol as in \cite{kowalski2010beyond}.
The mixing filters were room impulse responses simulated via the Roomsim toolbox \cite{campbell2005roomsim}, with a room size of dimension $3.55$ m $\times$ $4.45$ m $\times$ $2.5$ m, and with the same microphones and source configuration as in\cite{kowalski2010beyond}. 
%and in the SASSEC and SISEC evaluation campaigns \cite{vincent2007first, vincent20092008,5605570}.
The number of microphones was $M=2$, and the number of sources was varied in the range $3 \leq N \leq 6$.
The distances of the sources from the center of the microphone pairs was varied between $80$ cm and $1.2$ m. 
The mixing filters were generated with a reverberation time $RT_{60}$ of $250$ ms, 
and a microphone spacing of one meter.
%\revision{We also provide an experiment using Binaural Room Impulse Responses (BRIRs) \cite{hummersone2010dynamic}  captured in a large room of dimension $8.72$ m $\times$ $8.02$ m $\times$ $4.25$ m  with a  $RT_{60}$ of $890$ ms (room D in \cite{hummersone2010dynamic}).
%A Cortex (MK.2) Head and Torso Simulator (HATS), positioned at ($4.36$ m, $3.73$ m, $1.7$ m), and Genelec 8020A loudspeakers,
%were used to capture the responses.  
%The loudspeakers were placed around the HATS on an arc in the median plane with a 1.5 m radius and angles $45^{\circ}, 15^{\circ}, -10^{\circ}, -5^{\circ}$.
%}
For each case $N=3$ to $6$, ten mixtures where realized by convolving, for each mixture,  
$M$ mixing filters with $N$ music sources of the BSS Oracle dataset\footnote{available at \url{http://bass-db.gforge.inria.fr/bss_oracle}} \cite{Vin07b}
composed of $30$ music signals.
%\revision{i.e. with a similar protocol as in \cite{Vin07b,Vin09a,kowalski2010beyond,5466223,arberet2012tractable}.}
%discussion about noise
%We choose to not add additional noise to the mixture in order to only evaluate the source separation performance of the algorithms.
For all the constrained methods, we set $\epsilon = 10^{-4}$, and we vary the low-rank parameter from $\rindice=5$ to $\rindice=30$. 
%Note that ideally we would have set  $\epsilon = 0$ or used the proximity operator \eqref{eq_BP_LS} in the noise-free case, but 
%both approaches take an infinite number of iterations to reach convergence, and thus we need anyway to specify a tolerance.
We also compared our algorithm with the classical DUET method \cite{Yil04} as well as \SARABPDN\  \cite{arberet2013sparse} and the synthesis-$\ell_1$ minimization with 
wideband data-fidelity (BPDN-S) \cite{kowalski2010beyond,arberet2013sparse}.
%(i.e. the clustering step of DUET for mixing filters estimation is skipped and the source estimation step of DUET is initialised with the known mixing system $\mixop$), 
%\revision{as well as the results of the near-optimal 
%\footnote{
%Due to the non-orthogonality of the STFT, oracle masks as defined in \cite{Vin07b} (there are other definitions of the binary mask as in \cite{roman2011intelligibility}) have to be determined jointly in all TF points. As it is infeasible for realistic signals involving millions  of samples, we rather obtained near-optimal masks by minimizing the distortion on the target estimate in each TF point separately.
%} 
%binary mask \cite{Vin07b} which give an upper bound on the performance we can expect with any binary masking method such as DUET or others \cite{pedersen2008two,mandel2010model,sawada2011underdetermined}.}

The performance is evaluated for each source using the signal-to-distortion ratio (SDR),
%signal-to-interference ratio (SIR) and signal-to-artefact ratio (SAR), 
as defined in  \cite{gribonval2003proposals},
which indicates the overall quality of each estimated source compared to the target.
We then average this measure over all the sources and all the mixtures for each mixing condition.

%\subsection{Performance with respect to the number of sources}
%
%We evaluated the different methods in the case of a mixture of $4$ sources,
%we compare now the different methods for mixtures with $3 \leq N \leq 6$ sources and with $RT_{60}=250$ ms and $d=1$ m.

\subsection{Results}

\begin{figure}[!h]
  \centering 
      \centering \includegraphics[width= \myfigwidth]{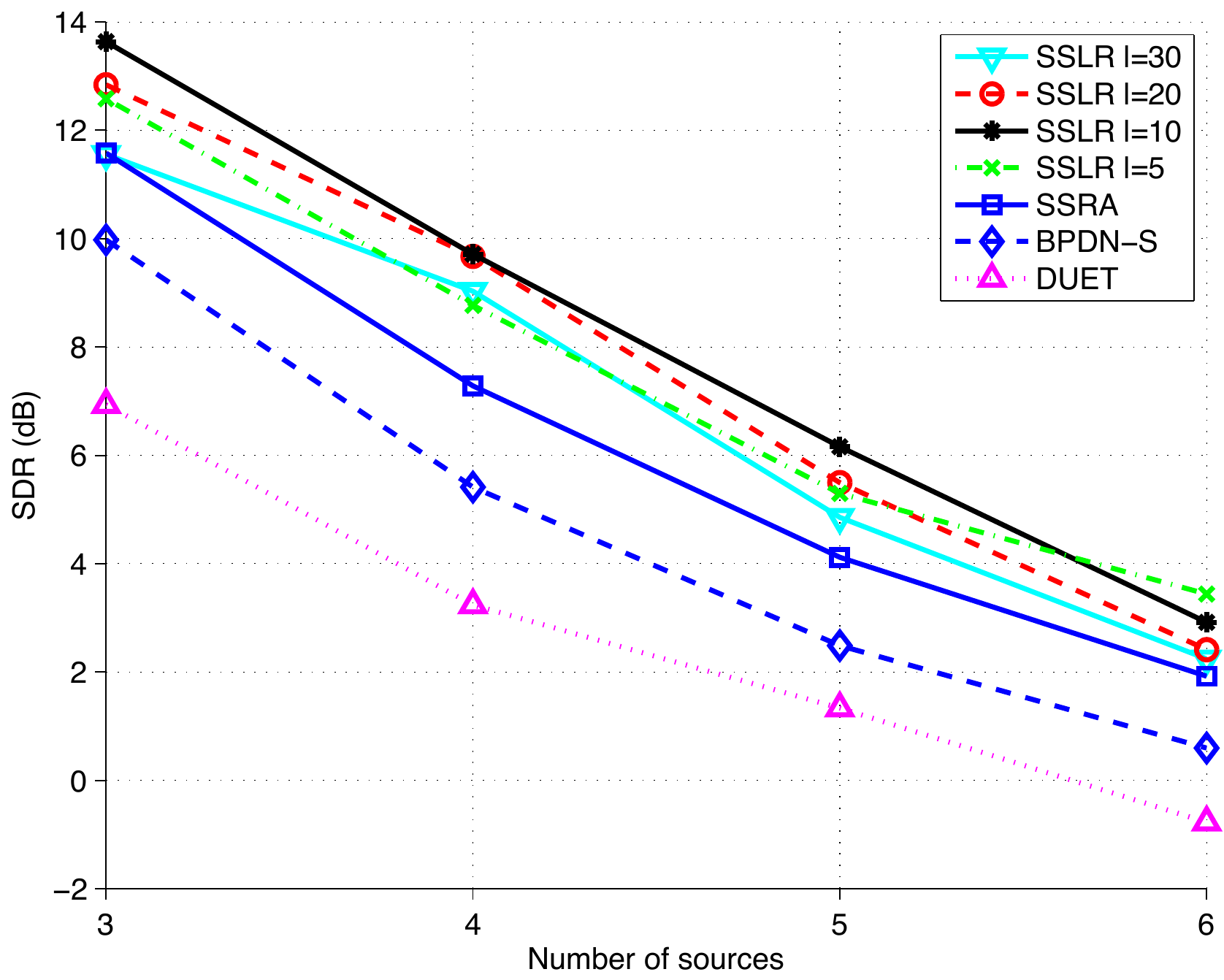}
      \caption{Variation of the average SDR as a function of the number $N$ of sources over music mixtures with reverberation time $RT_{60}=250$. \label{fig_expe3}}
\end{figure}

The results are depicted in Fig. \ref{fig_expe3}.
We can notice that the best performance is achieved with our proposed \SLRRA\ method with a maximal rank of $\rindice=10$. 
The improvement with respect to \SARABPDN\ %(which has the same formulation as \SLRRA\ but without the low-rank constraint) 
is about $2\pm1$ dB  in SDR depending of the number of sources. This shows the relevance of the low-rank constraint to model the source spectrograms. 
Moreover, all the other versions of \SLRRA, with other rank constraints $\rindice$, outperformed \SARABPDN, which indicates that the low-rank constraint does not degrade the performance even when $\rindice$ is not set optimally.

\section{Conclusion}

We proposed a novel algorithm for reverberant audio source separation, which exploits the structure of the sources via a (analysis) sparse
and low-rank prior on the source spectrograms.
The sources are estimated via an optimization algorithm derived from the ADMM proximal scheme and the Chambolle-Pock preconditioner.
The algorithm is also based on a reweighing analysis $\ell_1$ approach so as to increase the sparsity and a wideband data-fidelity term in a constrained form.
The results  on convolutive music mixtures show that the proposed method outperforms all of the tested methods 
with an improvement of $2\pm1$ dB of SDR over \SARABPDN, and $5\pm1.5$  dB over DUET.
%These results show how beneficial it is to add structural priors on the sources.
 %and that we can handle them using the framework of proximal splitting methods. 
%A natural extension of this work will then be to explore and incorporate other source structures and develop optimization algorithms that will handle them.
An extension of this work would be, in addition to the sources estimation, to estimate the mixing filters, possibly with an alternating optimization approach.

%\appendices
%\section{Appendix}
\appendix

\subsection{Proximity operators}

We derive the proximity operators for the functions
$f_1(\mathbf{s} ) = \|  \mathbf{s} \stftFrame  \|_{\weightMtx,1}$ and $f_2(\mathbf{s}) = i_{\mathcal{B}_{\ell_2}^{\epsilon}}(\mathbf{s})$,
and $f_3(\mathbf{s}) = i_{\mathcal{R}^{\rindice}}(\mathbf{s})$ introduced in section \ref{convexOptim}.

\begin{proposition}\emph{(Prox of $f_1( \cdot ) = \|\cdot \stftFrame  \|_{\weightMtx,1}$)}\label{prop_prox_f1}
Let $\tilde{\mathbf{z}} \in \mathbb{C}^{N \times \nbcoeff}$ and $\mathbf{z} \in \mathbb{R}^{N \times \nbsamples}$.
If $\stftFrame \in \mathbb{C}^{\nbsamples \times \nbcoeff}$ is a tight frame, i.e. $\stftFrame  \stftFrame^* = \framebound \Id$,
and $\weightMtx \in \mathbb{R}_+^{N \times \nbcoeff}$ is a matrix of positive weights  $\weightcoeff_{ij}$, 
 then 
\begin{equation}
\prox_{\|\cdot \stftFrame  \|_{\weightMtx,1}}({\mathbf{z}}) ={\mathbf{z}} + \framebound^{-1}( \prox_{\framebound \|\cdot  \|_{\weightMtx,1}} - \Id)(\mathbf{z} \stftFrame) \stftFrame^*, 
\end{equation}
with
\begin{equation}
\prox_{\framebound \| \cdot  \|_{\weightMtx,1}} (\tilde{\mathbf{z}})= (\prox_{\framebound \weightcoeff_{ij} | \cdot |} (\tilde{z}_{ij}) )_{1 \leq i \leq N, 1 \leq j \leq \nbcoeff}, 
\end{equation}
where $\prox_{\framebound \weightcoeff_{ij} | \cdot |}$ is the soft thresholding operator given
by $\prox_{\lambda |\cdot|}(z_i)  = \frac{z_i}{|z_i|}(|z_i| - \lambda)^{+}$ with $\lambda = \framebound \weightcoeff_{ij}$ and $(\cdot)^+ = \max(0,\cdot)$.
\end{proposition}
The proof of this proposition can be found in \cite{arberet2013sparse}.

 \begin{proposition}\emph{(Prox of $f_2( \cdot ) = i_{\mathcal{B}_{\ell_2}^{\epsilon}}(\cdot),$ i.e. $\mathcal{P}_{\mathcal{B}_{\ell_2}^{\epsilon}}(\cdot)$)}
\begin{align}
 \mathcal{P}_{\mathcal{B}_{\ell_2}^{\epsilon}}(\mathbf{z}) &= %\prox_{i_{\| \cdot - \mathbf{x}\|_2 \leq\epsilon}}(\mathbf{z}) \nonumber\\ 
 \mathbf{x} + \min(1,\epsilon/\| \mathbf{z}-\mathbf{x} \|_2)(\mathbf{z}-\mathbf{x}).
\end{align}
\end{proposition}

\begin{proposition}\emph{(Prox of $f_3(\cdot) = i_{\mathcal{R}^{\rindice}}(\cdot),$ i.e. $\mathcal{P}_{\mathcal{R}^{\rindice}}(\cdot)$)}
\begin{align}
\mathcal{P}_{\mathcal{R}^{\rindice}}(\mathbf{z}) & = \left( \LRProj( |  \stftFrameOp(\mathbf{z}_n) | ) \circ e^{i \angle  \stftFrameOp(\mathbf{z}_n)}  \right)_{1\leq n \leq N}, 
 \end{align}
\end{proposition}
\noindent with $e^{i \angle \mathbf{z}}: \mathbf{z} \mapsto \mathbf{y}=e^{i \angle \mathbf{z}}$ being the element wise phase such that $y_{nm} = e^{i \arg (z_{nm})}$, and
$\LRProj(\mathbf{z})$ being the projection onto the (non-convex) set $\LRset = \left\{\mathbf{z} : \rank(\mathbf{z})\leq \rindice \right\}$  of matrices having a rank less or equal than $\rindice$, which closed form solution, given by the Eckart-Young theorem \cite{eckart1936approximation} is: 
$\LRProj(\mathbf{z}) = \mathbf{u} \mySigma^{\rindice} \mathbf{v}^*$,
 where $\mathbf{z} = \mathbf{u} \mySigma \mathbf{v}^*$ is the singular value decomposition (SVD) of $\mathbf{z}$ and 
 $\mySigma$ is a diagonal matrix with non-increasing entries $\mySigmaEntry_{ii}$, and
 $\mySigmaEntry^{\rindice}_{ii} :=  \left\{ 
  \begin{array}{l l}
   \mySigmaEntry_{ii}  & \quad \text{if $i\leq \rindice$}\\
    0 & \quad \text{if $i > \rindice$.}
  \end{array} \right.$

Proof:
Let $\LRMagset$ be the set of complex matrices which element-wise magnitude is a low-rank matrix, i.e. $\LRMagset = \left\{\mathbf{z} : \rank(|\mathbf{z}|)\leq \rindice \right\}$ and  let $\LRMagProj(\mathbf{z}) = \argmin_{\mathbf{y}} \left\{ \| \mathbf{y} - \mathbf{z} \|_F: \mathbf{y} \in \LRMagset \right\}$ be the projection onto $\LRMagset$.
For any matrices $\mathbf{z}$ and $\mathbf{y}$, we have 
\begin{align}
\| \mathbf{y} - \mathbf{z} \|_F^2  &= \| |\mathbf{y}| \|_F^2 + \| |\mathbf{z}| \|_F^2 - 2 \tr \left( |\mathbf{z}|^{\transpose}\left(|\mathbf{y}|e^{i (\angle \mathbf{y}-\angle \mathbf{z})} \right) \right) \nonumber \\
& \geq \| |\mathbf{y}| - |\mathbf{z}| \|_F^2.  \label{eq:lowerboundFro}
\end{align}
 Inequality \eqref{eq:lowerboundFro} is an equality when $\angle \mathbf{y} = \angle \mathbf{z}$.
 Thus, if the phase of $\mathbf{y}$ is not constrained as in the set $\LRMagset$, the matrix $\mathbf{y}$ 
 minimizing $\| \mathbf{y} - \mathbf{z} \|_F$ is the one minimizing $ \| |\mathbf{y}| - |\mathbf{z}| \|_F^2$ with $\angle \mathbf{y} = \angle \mathbf{z}$.
 Then, $\LRMagProj(\mathbf{z}) = \argmin_{\mathbf{y}} \left\{ \| |\mathbf{y}| - |\mathbf{z}|  \|_F: |\mathbf{y}| \in \LRset, \angle \mathbf{y} = \angle \mathbf{z} \right\}  = \LRProj( | \mathbf{z} | ) \circ e^{i \angle \mathbf{z}}$.

\subsection{Chambolle-Pock preconditioner \cite{chambolle2011first} \label{sec_ChambollePock}}

The $\frstvariableA$-update step:
\begin{align}
\frstvariableA^{(\iter+1)}  & = 
\prox_{\gamma G}^{\compOpBig} (\thvariableA^{(\iter)} - \scvariableA^{(\iter)}) \nonumber \\
 & \eqdef \argmin_{\frstvariableA \in \mathcal{H}}  \gamma G(\frstvariableA) + \frac{1}{2} \|\compOpBig (\frstvariableA) - (\thvariableA^{(\iter)}-\scvariableA^{(\iter)}) \|^2 \label{eq_xupdateADMM}
 \end{align}
in the ADMM Algorithm \ref{algo_ADMM} is a  least squares problem including the linear operator $\compOpBig$ which computation necessitates inner iterations.
%Antonin Chambolle and Thomas Pock \cite{chambolle2011first} proposed a
The Chambolle-Pock preconditioner consists in adding, in the minimization \eqref{eq_xupdateADMM},   
the following term:
$\frac{1}{2} \inprod{(\frac{1}{\tau} - \frac{1}{\gamma} \compOpBig\compOpBig^*)(\frstvariableA-\frstvariableA^{(\iter)}), \frstvariableA-\frstvariableA^{(\iter)}}$,
with $\tau < \frac{\gamma}{\| \compOpBig \|^2}$.  
As a result the $\frstvariableA$-update  step becomes:
$$\frstvariableA^{(\iter+1)} = \prox_{\tau G} (\frstvariableA^{(\iter)} -\tau \compOpBig^* ( \bar{\frstvariableA}^{(\iter)})),$$
with $\bar{\frstvariableA}^{(\iter)} = \frac{1}{\gamma}(2 \scvariableA^{(\iter)} -\scvariableA^{(\iter-1)}).$

\bibliographystyle{IEEEtran}
\bibliography{article}

\end{document}